\documentclass[twocolumn,showpacs,preprintnumbers,amsmath,amssymb,superscriptaddress]{revtex4}

\usepackage{slashbox}
\usepackage{multirow}

\newcommand{\ket}[1]{{\left| #1 \right\rangle}}
\newcommand{\bra}[1]{{\left\langle #1 \right|}}
\begin{document}

\title{Secure quantum cryptographic network based on quantum key distribution}

\author{Sora Choi}\email{srchoi@math.snu.ac.kr}
\affiliation{Future Technology Research Division,
Electronics and Telecommunications Research Institute,
Daejeon 305-350, Korea}
\author{Soojoon Lee}\email{level@kias.re.kr}
\affiliation{School of Computational Science,
Korea Institute for Advanced Study, Seoul 130-722, Korea}
\author{Dong Pyo Chi}\email{dpchi@math.snu.ac.kr}
\affiliation{School of Mathematical Sciences,
Seoul National University, Seoul 151-742, Korea}
\date{\today}

\begin{abstract}
We present a protocol for quantum cryptographic network
consisting of a quantum network center and many users,
in which any pair of parties with members chosen from the whole users on request
can secure a quantum key distribution by help of the center.
The protocol is based on the quantum authentication scheme
given by Barnum~{\it et al.} [Proc. 43rd IEEE Symp. FOCS'02, p. 449 (2002)].
We show that
exploiting the quantum authentication scheme
the center can safely make two parties share
nearly perfect entangled states used
in the quantum key distribution.
This implies that
the quantum cryptographic network protocol is secure
against all kinds of eavesdropping.
\end{abstract}

\maketitle

%%%%%%%%%%%%%%%%%%%%%%%%%%%%%%%%%%%%%%%%%%%%%%%%%%%%%%%%%%%%%%%%%%%%%%
%%%                                                                %%%
%%%                          Introduction                          %%%
%%%                                                                %%%
%%%%%%%%%%%%%%%%%%%%%%%%%%%%%%%%%%%%%%%%%%%%%%%%%%%%%%%%%%%%%%%%%%%%%%
\section{Introduction}
%%%
%%%     quantum cryptographic protocols
%%%
Quantum key distribution (QKD)
has been considered
as a method for the perfectly secure communication
which can compensate for the incompleteness of the classical cryptography.
Since the advent of the first QKD protocol
presented by Bennett and Brassard~\cite{BB},
various kinds of
quantum cryptographic protocols
based on QKD schemes between two users~\cite{EK,BBM,B92}
have been proposed,
and the rigorous proofs of the security of the protocols
have considerably been studied~\cite{M,LC,SP,ILM,GLLJ,QT,KP,TKI}.
This implies that quantum cryptography has almost attained to the practical stage.

%%%
%%%     m-n QKD
%%%
The QKD protocol between two users can be
generalized into the communications between two parties, $A$ and $B$,
consisting of several members respectively~\cite{SG,CKC},
which use the quantum secret sharing protocol~\cite{HBB}.

%%%
%%%     quantum cryptographic network protocols
%%%
For the communication among a lot of users,
the theories on the quantum cryptographic network,
in which several protocols
are feasible on request,
have been suggested~\cite{BHM,Zeng,XLG}.
The quantum cryptographic network usually requires a quantum network center,
which connects any pair of parties by entangled states
so that the two parties can perform a quantum key distribution by help of the center.

Furthermore, employing the quantum network center,
one can reduce the number of quantum channels
used in quantum cryptographic network.
In other words, in order for any pair of $n$ users
to communicate with each other,
$n(n-1)/2$ quantum channels between any pairs of users are required
in the quantum cryptographic network without a quantum network center,
while only $n$ quantum channels between the center and users
are required
in the quantum cryptographic network with a quantum network center.
Therefore, the center's role can provide us with the efficient quantum cryptographic network
as well as the secure network.

%%%
%%%     our work
%%%
In this paper,
we present a protocol for the quantum cryptographic network,
which is based on the quantum message authentication scheme
presented by Barnum~{\it et~al.}~\cite{BCGST}.
In order to show that the quantum cryptographic network protocol is secure,
we first generalize the quantum authentication scheme between two persons into the scheme
between any pair of parties,
and then prove that
exploiting the quantum authentication scheme
the center can safely make any pair of parties share
nearly perfect entangled states used in the QKD
so that
the protocol is secure
against all kinds of eavesdropping.

%%%
%%%    organization of this paper
%%%
This paper is organized as follows.
In Sec.~\ref{sec:Preliminaries}
we briefly introduce the multipartite entangled states
used in the QKD between two parties
and their properties.
In Sec.~\ref{Sec:QKD3}
we present two protocols for the quantum cryptographic network.
In Sec.~\ref{QKD3:security}
we show that our protocols is secure against
any eavesdropping.
Finally, in Sec.~\ref{sec:Conclusions} we summarize our results.

%%%%%%%%%%%%%%%%%%%%%%%%%%%%%%%%%%%%%%%%%%%%%%%%%%%%%%%%%%%%%%%%%%%%%%
%%%                                                                %%%
%%%                   Preliminaries                                %%%
%%%                                                                %%%
%%%%%%%%%%%%%%%%%%%%%%%%%%%%%%%%%%%%%%%%%%%%%%%%%%%%%%%%%%%%%%%%%%%%%%
\section{Multipartite entangled states and their properties}\label{sec:Preliminaries}
For a positive integer $j$,
we define $\ket{\Phi_j^\pm}$ and $\ket{\Psi_j^\pm}$ by
\begin{align}
\ket{\Phi_j^\pm}
 &=\frac{1}{\sqrt{2}}\left(\ket{0^j}\pm \ket{1^j}\right), \nonumber\\
\ket{\Psi_j^\pm}
 &= \frac{1}{\sqrt{2}}\left(\ket{0^j}\pm \iota\ket{1^j}\right),\label{cat state1}
\end{align}
where $\ket{0}$ and $\ket{1}$ are the spin-up and the spin-down in the $z$-direction respectively,
and $\iota=\sqrt{-1}$.
Then
we can readily obtain the following decomposition relations~\cite{CKC}:
For positive integers $n$ and $m$ satisfying $n>m$,
\begin{align}
\ket{\Phi_n^\pm}
 & =\frac{1}{\sqrt{2}}\left(\ket{\Phi_{m}^+}\ket{\Phi_{n-m}^\pm} + \ket{\Phi_{m}^-}\ket{\Phi_{n-m}^\mp}\right)\nonumber\\
 &=\frac{1}{\sqrt{2}}\left(\ket{\Psi_{m}^+}\ket{\Psi_{n-m}^\mp} + \ket{\Psi_{m}^-}\ket{\Psi_{n-m}^\pm}\right),\nonumber\\
\ket{\Psi_n^\pm}
& =\frac{1}{\sqrt{2}}\left(\ket{\Phi_{m}^+}\ket{\Psi_{n-m}^\pm} + \ket{\Phi_{m}^-}\ket{\Psi_{n-m}^\mp}\right)\nonumber\\
& =\frac{1}{\sqrt{2}}\left(\ket{\Psi_{m}^+}\ket{\Phi_{n-m}^\pm} + \ket{\Psi_{m}^-}\ket{\Phi_{n-m}^\mp}\right).
\label{eq:relation}
\end{align}

We consider the case that each member of two parties $A$ and $B$,
consisting of $m$ and $n-m$ users respectively,
possesses one particle of an $n$-particle state $\ket{\Phi_n^\pm}$,
and that
each member measures one's own particle in the $x$- or $y$-direction.
For a party $P$ (with $l$ members),
let $\mathcal{Y}_{P}$ be the number (modulo~4) of members of $P$
who measure in the $y$-direction,
$\bar{\mathcal{Y}}_P = \left\lfloor\mathcal{Y}_{P}/{2}\right\rfloor$,
and
$\mathcal{M}_{P}$ the sum (modulo~2) of the measurement outcomes of all members in $P$.
Then we straightforwardly obtain the following properties from Eq.~(\ref{eq:relation}):
If $\mathcal{Y}_P$ is even, then
$\mathcal{M}_{P}\oplus \bar{\mathcal{Y}}_{P}$ is zero (or one)
when they share $\ket{\Phi_{l}^{+}}$ (or $\ket{\Phi_{l}^{-}}$),
where $\oplus$ is the addition modulo 2.
If $\mathcal{Y}_{P}$ is odd, then
$\mathcal{M}_{P}\oplus \bar{\mathcal{Y}}_{P}$ is zero (or one)
when $P$ has the state $\ket{\Psi_{l}^{+}}$ (or $\ket{\Psi_{l}^{-}}$).
\begin{table}
\begin{ruledtabular}
\begin{center}
\begin{tabular}[t]{c|c|cc|cc}
& & \multicolumn{2}{c}{A} & \multicolumn{2}{c}{B}
\\\hline
  &  $\mathcal{Y}_A \oplus \mathcal{Y}_B$ &
$\mathcal{Y}_A$ &
$\bar{\mathcal{Y}}_A\oplus\mathcal{M}_A$ &
$\mathcal{Y}_B$ &
$\bar{\mathcal{Y}}_B\oplus\mathcal{M}_B$\\
\hline
 \multirow{4}{16mm}{$\ket{\Phi_n^+}$ }&\multirow{4}{8mm}{even}
 &\multirow{2}{8mm}{even} & 0 & \multirow{2}{8mm}{even}  & 0 \\
 &  &  & 1 &   & 1\\
 \cline{3-6}
 & &\multirow{2}{8mm}{odd} & 0 & \multirow{2}{8mm}{odd}  & 1\\
 &     &      & 1   &                & 0\\ %\hline
  \end{tabular}
\end{center}
\end{ruledtabular}
\caption{
The relations between the measurement outcomes
of two parties $A$ and $B$.}\label{Table:Exencoding1}
\end{table}

\begin{table}
\begin{ruledtabular}
\begin{center}
\begin{tabular}[t]{c|cc|cc|cc}
&\multicolumn{2}{c}{C} & \multicolumn{2}{c}{A} &
\multicolumn{2}{c}{B}
\\\hline
  &  $\mathcal{Y}_C$ &
$\bar{\mathcal{Y}}_C\oplus \mathcal{M}_C$ &
$\mathcal{Y}_A$ &
$\bar{\mathcal{Y}}_A\oplus \mathcal{M}_A$ &
$\mathcal{Y}_B$ &
$\bar{\mathcal{Y}}_B\oplus \mathcal{M}_B$\\
\hline
 \multirow{8}{7mm}{$\ket{\Phi_n^+}$ } &\multirow{4}{4mm}{0}& 0
 &\multirow{2}{4mm}{even} & 0 (1) & \multirow{2}{4mm}{even}  & 0 (1) \\
 &  & 1 & & 0 (1) &   & 1 (0)\\
 & & 0 &\multirow{2}{4mm}{odd} & 0 (1) & \multirow{2}{4mm}{odd}  & 1 (0)\\
 & & 1 &     & 0 (1)   &                & 0 (1)\\ \cline{2-7}
 &\multirow{4}{4mm}{1}& 0
 &\multirow{2}{4mm}{even} & 0 (1) & \multirow{2}{4mm}{odd}  & 1 (0) \\
 &  & 1& & 0 (1) &   & 0 (1)\\
 & & 0 &\multirow{2}{4mm}{odd} & 0 (1) & \multirow{2}{4mm}{even}  & 1 (0) \\
 & &  1 &      & 0 (1)   &                & 0 (1)\\
 \end{tabular}
\end{center}
\end{ruledtabular}
\caption{The relations between the measurement outcomes of
three parties $A$, $B$, and $C$.}\label{Table:Exencoding2}
\end{table}
Hence, we get the relations between outcomes
of two parties and three parties as in the Table \ref{Table:Exencoding1} and
\ref{Table:Exencoding2}, respectively.
It follows from these relations that
not only the QKD between any kinds of two parties,
but also the QKD between two parties with a center's assistance
are feasible if they share the multi-particle entangled states
such as $\ket{\Phi_n^+}$~\cite{CKC}.

%%%%%%%%%%%%%%%%%%%%%%%%%%%%%%%%%%%%%%%%%%%%%%%%%%%%%%%%%%%%%%%%%%%%%%
%%%                                                                %%%
%%%           Protocols for quantum cryptographic network          %%%
%%%                                                                %%%
%%%%%%%%%%%%%%%%%%%%%%%%%%%%%%%%%%%%%%%%%%%%%%%%%%%%%%%%%%%%%%%%%%%%%%
\section{Protocols for quantum cryptographic network}\label{Sec:QKD3}
In this section,
we construct two slightly different protocols for quantum cryptographic network
according as the center has a memory
to store quantum information or does not have such a memory.
If the center has the memory,
then it can enhance the efficiency for the use of entangled states
by reducing the number of discarded entangled states.

Before presenting our protocol,
we briefly review
the results
of the quantum message authentication
presented by Barnum~{\it et~al.}~\cite{BCGST}:
The quantum authentication uses stabilizer codes
based on a normal rational curve in the projective geometry.
Barnum~{\it et~al.} proved that
the set of the codes forms a stabilizer purity testing code with error $\varepsilon=2r/(2^s+1)$,
where each code encodes $t=(r-1)s$ qubits into $u=rs$ qubits,
and also showed that
a secure quantum authentication scheme
can be obtained from the purity testing code.

%%%
%%%      Protocol 1: QKD network without memory
%%%
\subsection{Protocol 1: QKD network without memory}
We now present the QKD protocol between two users or two parties
which can securely be performed by help of the center's authentication for their states.
Throughout the paper,
we assume that $C$ is considered as one quantum network center which is always trustful,
and that $A$ and $B$ are two arbitrary parties
which want to be connected by
nearly perfect entangled states
in order to communicate each other.
The protocol is as follows:
\begin{itemize}\label{pro:authentication}
\item[(1)] Preprocessing: For each member $\mu$ in $A$ and $B$,
$C$ and $\mu$ agree on some stabilizer purity testing code $\{D_{\mathbf{k}(\mu)}\}$
and some private and random binary strings $\mathbf{k}(\mu)$, $\mathbf{x}(\mu)$, and $\mathbf{y}(\mu)$.
\item[(2)] $C$ prepares  $nt$ qubits in the state $\ket{\Phi_n^+}^{\otimes t}$.
We denote one qubit of $\ket{\Phi_n^+}$ which will be transmitted to the member $\mu$ by $\rho(\mu)$.
\item[(3)] Performing some specific unitary operations corresponding to $\mathbf{x}(\mu)$,
$C$ encrypts $\rho(\mu)^{\otimes t}\equiv\tilde{\rho}(\mu)$
as $\tau(\mu)$.
\item[(4)] For the code $D_{\mathbf{k}(\mu)}$ with syndrome $\mathbf{y}(\mu)$,
$C$ encodes $\tau(\mu)$ according to $D_{\mathbf{k}(\mu)}$ to produce $\sigma(\mu)$.
$C$ sends $\sigma(\mu)$ to the member $\mu$.
Let $\sigma'(\mu)$ be the state which $\mu$ receives.
\item[(5)] Each member $\mu$ measures
the syndrome $\mathbf{y}'(\mu)$ of the code $D_{\mathbf{k}(\mu)}$ on $\sigma'(\mu)$.
Each member $\mu$ compares $\mathbf{y}(\mu)$ to $\mathbf{y}'(\mu)$
and aborts if any error is detected.
\item[(6)] According to $D_{\mathbf{k}(\mu)}$,
each member $\mu$ decodes $\sigma'(\mu)$ to obtain $\tau'(\mu)$,
and decrypts $\tau'(\mu)$ using $\mathbf{x}(\mu)$,
and then obtains a $t$-qubit state, $\tilde\rho'(\mu)$.
\item[(7)] $\mu$ randomly performs a measurement
on each qubit of one's own state
in the $x$- or $y$-direction.

After each member individually carries out the above steps,
each member in two parties takes $t$ bits as measurement results.
Thus, in order to obtain sufficiently many bit strings,
$C$ and the two parties repeat the above steps
as many times as necessary.

\item[(8)] For all shared qubits,
each member in two parties publicly announces the used directions,
but not the obtained results.
Then the two parties, $A$ and $B$, obtain
$\mathcal{Y}_A$ and $\mathcal{Y}_B$.
If $\mathcal{Y}_A\oplus\mathcal{Y}_B$ is odd then they discard the keys.
Otherwise, $A$ and $B$ continue the next step.

\item[(9)]
A collector in $A$ ($B$),
gathers the outcomes to obtain $\mathcal{M}_A$ ($\mathcal{M}_B$).

\item[(10)] Two parties $A$ and $B$ have a public discussion
on a random subset of the obtained bits,
which is used as the test bits,
in order to detect an error which may occur in the previous procedure.

If the parties find an error in this step,
all shared keys are discarded, and
they go back to Step~(1).
Otherwise, they obtain a final key string.
\end{itemize}
We remark that the steps~(9) and (10) are unnecessary
when both of two parties $A$ and $B$ consist of only one member.

We now review a method to obtain $\mathcal{M}_A$ (or $\mathcal{M}_B$)
presented in \cite{CKC}:
We first assume that all members are ordered
and that the first member is the collector.
This order need to be arbitrary.
The collector chooses a random bit,
which we will express as `$R$',
adds it to his outcome modulo 2,
and transfers the result to the second member.
The second member transfers the next member the outcome plus the received one modulo 2.
All members continue this procedure
until the collector receives $\mathcal{M}_A\oplus R$ (or $\mathcal{M}_B\oplus R$).
Then the collector obtains $\mathcal{M}_A$ (or $\mathcal{M}_B$),
which is $\mathcal{M}_A\oplus R\oplus R$ (or $\mathcal{M}_B\oplus R\oplus R$).
However, anyone cannot know $\mathcal{M}_A$ (or $\mathcal{M}_B$)
without acquiring the results of all members.

%%%
%%%      Protocol 2: QKD network with memory
%%%
\subsection{Protocol 2: QKD network with memory}
In here,
we introduce the protocol
in which the center does not only authenticate quantum states
but also joins in the procedure of the QKD scheme
so that the entangled states employed in QKD can efficiently be used.
The protocol can be obtained from modifying several steps in Protocol~1 as follows:
\begin{itemize}
\item[(2$'$)] $C$ prepares  $(n+1)t$ qubits in the state $\ket{\Phi_{n+1}^+}^{\otimes t}$.
$C$ stores $t$-qubit state $\rho(C)^{\otimes t}\equiv\tilde\rho(C)$
in his memory
where $\rho(C)$ is one qubit of $\ket{\Phi_{n+1}^+}$.
\item[(7$'$)] Each member in the
two parties, except the center $C$, randomly performs a measurement
on each one qubit in the $x$- or $y$-direction.
\item[(8$'$)]
\begin{enumerate}
\item[(a)] Each member in the two parties,
except the center $C$, publicly announces the used directions,
but not the obtained results.
Then $A$, $B$, and $C$ obtain $\mathcal{Y}_A$ and $\mathcal{Y}_B$.
\item[(b)] According to the values of $\mathcal{Y}_A$ and $\mathcal{Y}_B$,
$C$ performs a measurement on his particle
which is in the state $\rho(C)$
so that $\bar{\mathcal{Y}}_A\oplus\mathcal{M}_A$ and $\bar{\mathcal{Y}}_B\oplus\mathcal{M}_B$
have a correlation or an anti-correlation.
\end{enumerate}
\item [(9$'$)] Two parties, $A$ and $B$, collect the outcomes
to obtain $\mathcal{M}_A$ and $\mathcal{M}_B$ respectively,
as in the previous protocols.
Then the center $C$ reveals the obtained results.
\end{itemize}
We remark that
the entangled states are used
in Protocol~2 more efficiently than in Protocol~1
since there are cases that the keys are discarded in Step~(8) of Protocol~1
while there is not such a case in Step~(8$'$) of Protocol~2.

%%%%%%%%%%%%%%%%%%%%%%%%%%%%%%%%%%%%%%%%%%%%%%%%%%%%%%%%%%%%%%%%%%%%%%
%%%                                                                %%%
%%%        Proof of security of quantum cryptographic network      %%%
%%%                                                                %%%
%%%%%%%%%%%%%%%%%%%%%%%%%%%%%%%%%%%%%%%%%%%%%%%%%%%%%%%%%%%%%%%%%%%%%%
\section{Proof of security of quantum cryptographic network}\label{QKD3:security}
In this section, we are going to prove the security of the protocols presented in this paper.
We first note that
if, in Step~(2$'$) of Protocol~2,
$C$ measures each qubit of $\tilde{\rho}(C)$ in the $x$- or $y$-direction
instead of storing it
then the protocol is essentially equivalent to Protocol~1,
and that the order of the measurement does not affect the security of the protocol.
Thus, it suffices to prove the security of Protocol~2.

The secure quantum authentication scheme in~\cite{BCGST}
ensures that
for any $t$-qubit state $\tilde{\rho}$,
when $\tilde{\rho}$ is transmitted to a member $\mu$ %of the parties
as $\tilde{\rho}'$,
the fidelity $F$ of $\tilde{\rho}$ and $\tilde{\rho}'$ is not less than $1-\varepsilon_0$
for sufficiently small $\varepsilon_0>0$,
where the fidelity $F$ of $X$ and $Y$ is defined by
\begin{equation}
F(X,Y)=\mathrm{tr}\left(\sqrt{{X}^{1/2} Y {X}^{1/2}}\right)^2.
\label{eq:fidelity}
\end{equation}
Here, $\tilde{\rho}'$ can be considered as $\Lambda_\mu(\tilde{\rho})$
for some quantum channel $\Lambda_{\mu}$.
Thus,
if the transmitted state is not discarded, then
for any $t$-qubit $\tilde{\rho}$
we can obtain the inequality
\begin{equation}
F(\tilde{\rho},\tilde{\rho}')=F(\tilde{\rho},\Lambda_{\mu}(\tilde{\rho})) \ge 1-\varepsilon_0.
\label{eq:fidelity1}
\end{equation}

For the detailed proof,
we present a remark on a property of the fidelity and the quantum channel
presented in Lemma~3 of Ref.~\cite{LCC}:
Let $\mathcal{E}$ be a quantum operation on
a $d$-dimensional quantum system $\mathcal{H}_A$,
and $\ket{\Psi}\in \mathcal{H}_A\otimes \mathcal{H}_R$ a purification of
a state $\rho_A$ on $\mathcal{H}_A$,
where $\mathcal{H}_R$ is a reference system
such that $\mathrm{tr}_R(\ket{\Psi}\bra{\Psi})=\rho_A$.
Suppose that there is $\varepsilon > 0$ such that
\begin{equation}
\bra{\psi}\mathcal{E}(\ket{\psi}\bra{\psi})\ket{\psi}\ge 1-\varepsilon
\label{eq:support}
\end{equation}
for all $\ket{\psi}$ in the support of $\rho_A$.
Then
\begin{align}
\bra{\Psi}&\left[\left(\mathcal{E}\otimes\mathcal{I}_R\right)
\left(\ket{\Psi}\bra{\Psi}\right)\right]\ket{\Psi} \nonumber\\
&\ge 1-\left(1+d_0\cdot\max_{j\neq k}\{p_jp_k\}\right)\varepsilon,
\label{eq:entanglement_fidelity}
\end{align}
where $d_0$ is the Schmidt number of $\ket{\Psi}$
and $\sqrt{p_j}$ are the Schmidt coefficients of $\ket{\Psi}$
with respect to the bipartite quantum system
$\mathcal{H}_A\otimes \mathcal{H}_R$.

Since for any member $\mu$
there exists a sufficiently small $\varepsilon_1>0$ such that
\begin{equation}
F\left(\ket{\psi}\bra{\psi},\Lambda_{\mu}(\ket{\psi}\bra{\psi})\right) \ge 1- \varepsilon_1
\label{eq:fidelity2}
\end{equation}
for all $t$-qubit state $\ket{\psi}$ by the inequality (\ref{eq:fidelity1}) and
\begin{equation}
\ket{\Phi_{n+1}^+}^{\otimes t}=\frac{1}{\sqrt{2^t}}\sum_{j=0}^{2^t-1}\ket{j}_\mu\ket{j^{n}}_C,
\label{eq:Phi}
\end{equation}
applying the above remark to our situation,
we obtain the following property:
\begin{equation}
F\left(\Phi,\left(\Lambda_{\mu}\otimes\mathcal{I}_C\right)
\left(\Phi\right)\right)\ge 1-\left(1+\frac{1}{2^t}\right)\varepsilon_1
\label{eq:entanglement_fidelity2}
\end{equation}
where $\Phi=\left(\ket{\Phi_{n+1}^+}\bra{\Phi_{n+1}^+}\right)^{\otimes t}$.

As we discuss the details in Appendix,
$\bigotimes_\mu\Lambda_{\mu}$ is also a nearly perfect quantum channel
and hence
there is a sufficiently small $\varepsilon_2>0$ such that
\begin{equation}
F\left(\ket{\xi}\bra{\xi},\bigotimes_\mu\Lambda_{\mu}(\ket{\xi}\bra{\xi})\right)\ge 1-\varepsilon_2
\label{eq:fidelity_bigotimes}
\end{equation}
for any $nt$-qubit pure state $\ket{\xi}$.
Thus, since it follows from Eq.~(\ref{eq:Phi}) that
\begin{equation}
\ket{\Phi_{n+1}^+}^{\otimes t}=\frac{1}{\sqrt{2^t}}\sum_{j=0}^{2^t-1}\ket{j^n}_{AB}\ket{j}_C,
\end{equation}
when the center transmit $nt$ qubits of the states to all members,
we can obtain the almost same inequality:
\begin{equation}
F\left(\Phi,\left(\bigotimes_\mu\Lambda_{\mu}\otimes\mathcal{I}_C\right)
\left(\Phi\right)\right)\ge 1-\left(1+\frac{1}{2^t}\right)\varepsilon_2.
\label{eq:entanglement_fidelity3}
\end{equation}
%for the sufficiently small $\varepsilon_1>0$.
Therefore, after executing Step~(6),
the parties with the center share
the nearly perfect entangled states.

We now need a result of Lo and Chau~\cite{LC}
that if the center $C$ and every members %$\mu$'s
share a state having a fidelity exponentially close to 1
with $\ket{\Phi_{n}^{+}}^{\otimes t}$
then Eve's mutual information with the key is at most
exponentially small.
From this result,
we can directly notice that Protocol~2 is secure
if there are no dishonest members in the parties.

%%%
%%%     Dishonest members
%%%
For the analysis of the case that dishonest members exist in the parties,
we use the investigation presented in~\cite{CKC}.
By a classical probability estimate about the random sampling tests~\cite{SP},
if there are some errors in the procedure then
two parties can detect an error with high probability
from sufficiently many test bits.
We remark that
each member can have an effect on the protocol
just at the moment of announcing the basis or giving the information on the outcomes,
and that the test bits are chosen
after the directions are announced
and the information on the outcomes is transferred to other members.
Therefore,
since the test bits are randomly chosen,
the dishonest members cannot escape the detection, that is,
they cannot prevent the others from obtaining the correct keys
without being detected.

%%%%%%%%%%%%%%%%%%%%%%%%%%%%%%%%%%%%%%%%%%%%%%%%%%%%%%%%%%%%%%%%%%%%%%
%%%                                                                %%%
%%%                      Conclusions                               %%%
%%%                                                                %%%
%%%%%%%%%%%%%%%%%%%%%%%%%%%%%%%%%%%%%%%%%%%%%%%%%%%%%%%%%%%%%%%%%%%%%%
\section{Summary}\label{sec:Conclusions}
In this paper,
we have presented a protocol for quantum cryptographic network
consisting of a quantum network center and many users,
in which any pair of parties with some members
can secure a quantum key distribution by help of the center.
We have shown that
exploiting the quantum authentication scheme
the center can safely make two parties share
nearly perfect entangled states
used in the QKD on request.
We have also shown that
the quantum cryptographic network protocol is secure
against all kinds of eavesdropping.

%%%%%%%%%%%%%%%%%%%%%%%%%%%%%%%%%%%%%%%%%%%%%%%%%%%%%%%%%%%%%%%%%%%%%%
%%%                                                                %%%
%%%                     Acknowledgments                            %%%
%%%                                                                %%%
%%%%%%%%%%%%%%%%%%%%%%%%%%%%%%%%%%%%%%%%%%%%%%%%%%%%%%%%%%%%%%%%%%%%%%
\acknowledgments{
S.C. acknowledges the support from Ministry of Planning and Budget,
S.L. from a KIAS Research Fund (No.~02-0140-001),
and D.P.C. from a Korea Research Project (No.~M1-0326-08-0002-03-B51-08-002-12)
funded by the Korean Ministry of Science and Technology.
S.C. would like to thank Prof. V.~Bu\v{z}ek for helpful advices,
and
S.L. would like to thank Prof. Jaewan Kim and Dr. Sangchul Oh in KIAS
and Prof. Jinhyoung Lee in Hanyang University for useful discussions.
}

%%%%%%%%%%%%%%%%%%%%%%%%%%%%%%%%%%%%%%%%%%%%%%%%%%%%%%%%%%%%%%%%%%%%%%
%%%                                                                %%%
%%%                     Appendix                                   %%%
%%%                                                                %%%
%%%%%%%%%%%%%%%%%%%%%%%%%%%%%%%%%%%%%%%%%%%%%%%%%%%%%%%%%%%%%%%%%%%%%%
\appendix*
\section{}
In this appendix,
we show that
$\bigotimes_\mu\Lambda_{\mu}$ is also a nearly perfect quantum channel
and that
there is a sufficiently small $\varepsilon_2>0$ such that
\begin{equation}
F\left(\ket{\xi}\bra{\xi},\bigotimes_\mu\Lambda_{\mu}\left(\ket{\xi}\bra{\xi}\right)\right)\ge 1-\varepsilon_2
\label{eq:fidelity_bigotimes_App}
\end{equation}
for any $nt$-qubit pure state $\ket{\xi}$.

First we note that
the last term of the inequality~(\ref{eq:entanglement_fidelity})
is greater than or equal to
$1-\left(1+d/4\right)\varepsilon$
since $d_0p_jp_k\le d/4$ for all $j\neq k$.
Thus, in the property of the fidelity and the quantum channel
presented in Lemma~3 of Ref.~\cite{LCC},
if there is $\varepsilon > 0$ such that
\begin{equation}
\bra{\psi}\mathcal{E}(\ket{\psi}\bra{\psi})\ket{\psi}\ge 1-\varepsilon
\label{eq:support_App}
\end{equation}
for all $\ket{\psi}\in \mathcal{H}_A$,
then
\begin{align}
\bra{\Phi}\left[\left(\mathcal{E}\otimes\mathcal{I}_R\right)
\left(\ket{\Phi}\bra{\Phi}\right)\right]\ket{\Phi}
\ge 1-\left(1+\frac{d}{4}\right)\varepsilon,
\label{eq:entanglement_fidelity_App}
\end{align}
for all $\ket{\Phi}\in \mathcal{H}_A \otimes \mathcal{H}_R$.
Since
the square root of the fidelity $F$ is doubly concave~\cite{FG},
that is, for $\sum_j\lambda_j=1$
\begin{equation}
\sqrt{F\left(\sum_j\lambda_j\rho_j,\lambda_j\rho'_j\right)}
\ge \sum_j\lambda_j\sqrt{F\left(\rho_j,\rho'_j\right)},
\label{eq:double_concavity}
\end{equation}
it follows from the inequality~(\ref{eq:entanglement_fidelity_App}) that
if the inequality~(\ref{eq:support_App}) holds then
for all density matrices $\rho_{AR}$ on $\mathcal{H}_A \otimes \mathcal{H}_R$
\begin{align}
F\left(\rho_{AR},\left(\mathcal{E}\otimes\mathcal{I}_R\right)
\left(\rho_{AR}\right)\right) %\nonumber\\
\ge 1-\left(1+\frac{d}{4}\right)\varepsilon.
\label{eq:entanglement_fidelity_App2}
\end{align}
Hence, we obtain from the inequalities~(\ref{eq:entanglement_fidelity2})
and (\ref{eq:entanglement_fidelity_App2}) that
\begin{equation}
F\left(\Phi,\left(\Lambda_{\mu}\otimes\mathcal{I}_C\right)
\left(\Phi\right)\right)\ge 1-\left(1+2^{t-2}\right)\varepsilon_1
\label{eq:entanglement_fidelity_App3}
\end{equation}
and that
\begin{align}
F&\left(\left(\mathcal{I}_\mu\otimes
\Lambda_{\mu'}\otimes \mathcal{I}_C\right)\left(\Phi\right),
\left(\Lambda_{\mu}\otimes\Lambda_{\mu'}\otimes\mathcal{I}_C\right)
\left(\Phi\right)\right)\nonumber\\
&=F\left(\rho,
\left(\Lambda_{\mu}\otimes\mathcal{I}_{\mu' C}\right)
\left(\rho\right)\right)\nonumber\\
&\ge 1-\left(1+2^{t-2}\right)\varepsilon_1
\label{eq:entanglement_fidelity_App4}
\end{align}
for $\mu\neq\mu'$,
where $\rho=\left(\Lambda_{\mu'}\otimes \mathcal{I}_{\mu C}\right)\left(\Phi\right)$.

We now present a simple remark on a relation
between the fidelity and a distance of density matrices, $D$,
which is defined by $D(\varrho,\varrho')=\mathrm{tr}|\varrho-\varrho'|/2$~\cite{FG}:
For any density matrices $\varrho$ and $\varrho'$,
the following inequalities hold.
\begin{equation}
1-\sqrt{F(\varrho,\varrho')}\le D(\varrho,\varrho') \le \sqrt{1-F(\varrho,\varrho')}.
\label{eq:fidelity_distance}
\end{equation}
Then it follows from the inequality~(\ref{eq:fidelity_distance})
and the triangle inequality of $D$
that
for density matrices $\varrho$, $\varrho'$, and $\varrho''$
\begin{align}
1-\sqrt{F(\varrho,\varrho'')}
\le& D(\varrho,\varrho'')
\le D(\varrho,\varrho')+D(\varrho',\varrho'')\nonumber\\
\le& \sqrt{1-F(\varrho,\varrho')}+\sqrt{1-F(\varrho',\varrho'')}.
\label{eq:fideilty_triangle}
\end{align}

By virtue of the inequalities~(\ref{eq:entanglement_fidelity_App3}),
(\ref{eq:entanglement_fidelity_App4}), and (\ref{eq:fideilty_triangle}),
we obtain the followings:
for any $nt$-qubit pure state $\ket{\xi}$,
\begin{align}
\sqrt{F\left(\ket{\xi}\bra{\xi},\bigotimes_\mu\Lambda_{\mu}(\ket{\xi}\bra{\xi})\right)}
 \ge 1-n\sqrt{(1+2^{t-2})\varepsilon_1}.
\label{eq:fidelity_bigotimes_App2}
\end{align}
Since $\varepsilon_1$ is independent on $n$ and $t$,
the proof is completed.
Furthermore, it is clear that
the result similar to the inequality~(\ref{eq:entanglement_fidelity3})
can directly be obtained in the same way.

Considering the Bures metric $d_B$~\cite{Bures,UJ}
defined by
\begin{equation}
d_B(\rho,\rho')=\sqrt{2-2F(\rho,\rho')^{1/2}},
\label{eq:rho_rho'}
\end{equation}
since $d_B$ satisfies the triangle inequality
we can actually obtain an inequality tighter than the inequality~(\ref{eq:fideilty_triangle}):
For density matrices $\varrho$, $\varrho'$, and $\varrho''$,
\begin{align}
\sqrt{1-F(\varrho,\varrho'')^{1/2}}
\le&\sqrt{1-F(\varrho,\varrho')^{1/2}}\nonumber\\
&+\sqrt{1-F(\varrho',\varrho'')^{1/2}}.
\label{eq:fideilty_triangle2}
\end{align}
Therefore, from the inequality~(\ref{eq:fideilty_triangle2}),
we could also obtain an inequality tighter than
the inequality~(\ref{eq:fidelity_bigotimes_App2}).

%%%%%%%%%%%%%%%%%%%%%%%%%%%%%%%%%%%%%%%%%%%%%%%%%%%%%%%%%%%%%%%%%%%%%%
%%%                                                                %%%
%%%                 References                                     %%%
%%%                                                                %%%
%%%%%%%%%%%%%%%%%%%%%%%%%%%%%%%%%%%%%%%%%%%%%%%%%%%%%%%%%%%%%%%%%%%%%%

\end{document}